\documentclass[%
superscriptaddress,
prb
 amsmath,amssymb,
 aps,
floatfix,
twocolumn,
]{revtex4-1}

\usepackage{graphicx}
\usepackage{dcolumn}
\usepackage{bm}
\usepackage{colortbl}
\usepackage{amsmath} 

\usepackage{cleveref}

\begin{document}


\title{Relaxing graphene plasmon excitation constraints through the use of an epsilon-near-zero substrate}

\author{Vinicius Tadin Alvarenga}
\email{vinicius.t.alvarenga@gmail.com}

\author{D. A. Bahamon}

\affiliation{MackGraphe, Mackenzie Presbyterian Institute, Sao Paulo, SP 01302-907, Brazil}
\affiliation{School of Engineering, Mackenzie Presbyterian University, Sao Paulo, SP 01302-907, Brazil}

\author{Nuno M. R. Peres}
\affiliation{Department and Centre of Physics, University of Minho,
Campus of Gualtar, 4710-057, Braga, Portugal}
\affiliation{International Iberian Nanotechnology Laboratory (INL), Braga, Portugal}

\author{Christiano J. S. de Matos}
\email{cjsdematos@mackenzie.br}

\affiliation{MackGraphe, Mackenzie Presbyterian Institute, Sao Paulo, SP 01302-907, Brazil}
\affiliation{School of Engineering, Mackenzie Presbyterian University, Sao Paulo, SP 01302-907, Brazil}

\date{\today}

\begin{abstract}

Graphene plasmons have attracted significant attention due to their tunability, potentially long propagation lengths and ultracompact wavelengths. However, the latter characteristic imposes challenges to light-plasmon coupling in practical applications, generally requiring sophisticated coupling setups, extremely high doping levels and/or graphene nanostructuting close to the resolution limit of current lithography techniques. Here, we propose and theoretically demonstrate a method for alleviating such a technological strain through the use of a practical substrate whose low and negative dielectric function naturally enlarges the graphene polariton wavelength to more manageable levels. We consider silicon carbide (SiC), as it exhibits a dielectric function whose real part is between -1 and 0, while its imaginary part remains lower than 0.05, in the 951 to 970 cm$^{-1}$ mid-infrared spectral range. Our calculations show hybridization with the substrate's phonon polariton, resulting in a polariton wavelenth that is an order of magnitude longer than obtained with a silicon dioxide substrate, while the propagation length increases by the same amount. 
\end{abstract}

\pacs{Valid PACS appear here}
\maketitle


\section{Introduction}
Plasmons, the collective oscillation of free electrons in a conductive material, have attracted great attention in the field of nanophotonics due to its high confinement of electromagnetic energy, achieving subdifractional  spatial extensions\cite{abajoch,maier}, which allows for the construction of photonic devices at the nanometer scale. The applications range from biosensing\cite{bio1,bio2,bio3} to enhanced spectroscopy techniques such as SERS\cite{sers1,sers2,sers3}, TERS\cite{ters1,ters2,ters3,ters4} and SNOM\cite{snom1,snom2,snom3,snom5,snom5,snom6}.  

Graphene, a two-dimensional material consisting of a single layer of carbon atoms in a hexagonal lattice\cite{geim-nov-2004}, has emerged as a promising material for plasmonic applications due to its high electronic mobility and null bandgap, which allows for high Fermi level tunability and significant changes on the optical conductivity\cite{koppens,pol-2d}. Since the plasmon dispersion is dependent on the electronic density, graphene plasmons can be actively controlled more easily than those in conventional conductors\cite{fang}. Also, graphene can be transfered to a range of dielectric surfaces, allowing for the integration of graphene-based plasmonic devices with photonic devices, such as silicon-based waveguides\cite{livro-nuno}. 

The plasmon wavevector in graphene tends to be up to 2 orders of magnitude larger than that of free space radiation in the mid-infrared range. However, efficient radiation-plasmon coupling, for the excitation of surface plasmon polaritons (SPPs), requires circumventing the large wavevector mismatch. For bridging such a high mismatch, finding a suitable high-index dielectric material for prism coupling\cite{otto,kre} becomes impractical, while  diffraction grating coupling has required structures with periodicity of tens of nanometers\cite{grating1,grating2,grating3}, posing a strain on the device fabrication step. Alternatively, localized surface plasmons (LSPs) have been excited in graphene patterned into nanostructures such as nanoribbons or nanodisks\cite{ju,Rodrigo,fang2}. Again, due to the large wavevector mismatch, the features in these structures need to be in the tens to hundreds of nanometers range, for operation in the mid-infrared\cite{Rodrigo,fang,vasic}. Even though such structures have been fabricated and used to excite graphene plasmons, the fabrication of such devices is challenging due to their size being close to the spatial resolution limit of current nanofabrication techniques\cite{nikitin}. 

The plasmon dispersion in graphene is dependent on the dielectric function of the adjacent media. Therefore, the  choice of the substrate plays a key role in the  wavevector matching for plasmon excitation. In particular a decrease in the magnitude of the plasmon wavevector is expected for a substrate with a low permittivity\cite{abajoch,livro-nuno}. The plasmon wavevector is proportional to the sum of the dielectric functions of the substrate and superstrate.
If we consider the latter to be air, meaning a dielectric function equal to 1, the optimal choice of substrate would be one with dielectric function of -1. However, a common drawback of materials with a negative dielectric function is that the imaginary part of the dielectric function cannot be neglected, leading to plasmon losses.

Epsilon-near-zero (ENZ) materials have received considerable recent attention for a number of interesting properties, including increased nonlinear optical susceptibilities\cite{Alam}, which can be used to enhance nonlinear effects such as second harmonic generation\cite{pilar}, and wavelength enlargement, which allows light tunneling through narrow channels\cite{silveirinha}. As in plasmonics, ENZ applications also require a low imaginary dielectric function component at the spectral range where the real dielectric function is low. As a consequence, ENZ materials are also good candidates for plasmonic applications. Indeed, they have been studied in the field of plasmonics for being capable of directly coupling SPPs to free-space light\cite{traviss}. 

Silicon carbide is an ENZ material in the mid-infrared due to its optical phonons\cite{tiwald,engheta1,engheta2}. The range between the transverse optical phonon (TO), at 797 cm$^{-1}$, and the longitudinal optical phonon (LO), at 970 cm$^{-1}$, called the Reststrahlen band, exhibits a real negative dielectric function, with the function crossing zero at both these wavenumbers.\cite{caldwell,koch}. In particular, a low imaginary part of epsilon is obtained near the LO phonon wavenumber. Within the Reststrahlen band, the material also supports surface phonon polaritons (SPhP)\cite{koch,huber, sic1,sic2}. It has been shown that when graphene plasmons are excited within the Reststrahlen band of a SiC substrate, hybrid plasmon-phonon modes arise, which maintain the characteristics of both excitations, exhibiting a longer wavelength than a pure SPP and keeping losses low\cite{hy1,hy2,hy3,sic3,sic4}. The use of SiC is also convenient since graphene can be epitaxially grown directly onto this substrate\cite{epi1,epi2,epi3,epi4,epi5}. Other polar dielectrics have also been studied as substrates for graphene plasmonics due to their ENZ nature\cite{hy1,hy2,hy3}, however, ENZ substrates have not been used for increasing the graphene plasmon wavelength with the purpose of reducing the wavevector mismatch for optical excitation.
In this work, we propose the use of SiC as a graphene substrate for mitigating this mismatch in the mid-infrared. Our results show an increase from hundreds of nanometers to a few micrometers in polariton wavelength on SiC when compared with a SiO$_2$ substrate. This simplifies the polaritonic excitation while keeping the polariton modal volume 3 orders of magnitude smaller than that of free-space radiation. Although there is a loss of confinement, these polaritonic waves become easier to excite, and the fabrication of polariton-based devices becomes easier. The conditions for this trade-off can be adjusted by tailoring the size of the structures used for excitation and controlling the graphene doping. 

\begin{figure}[t]
    \centering
    \includegraphics[scale=0.28]{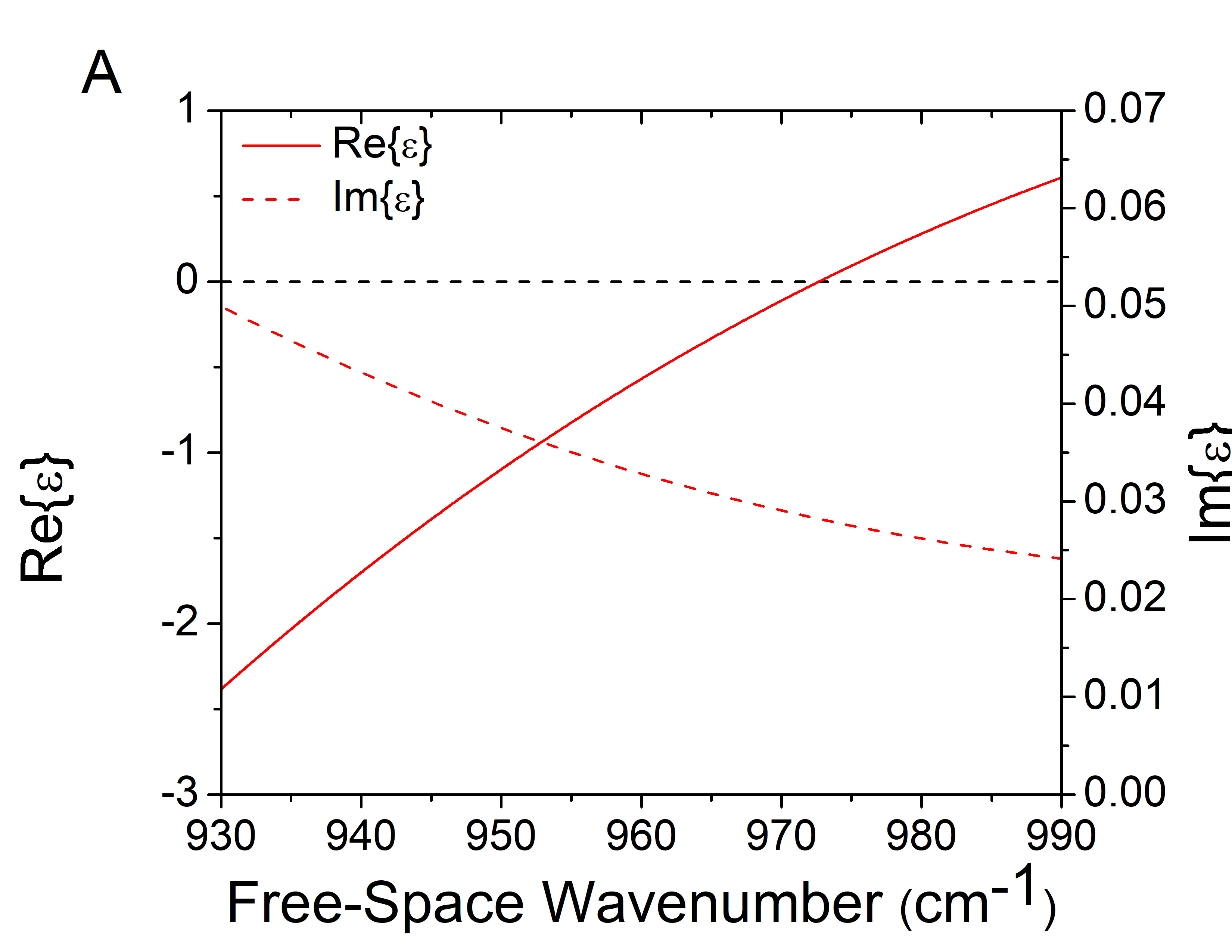}
    \includegraphics[scale=0.28]{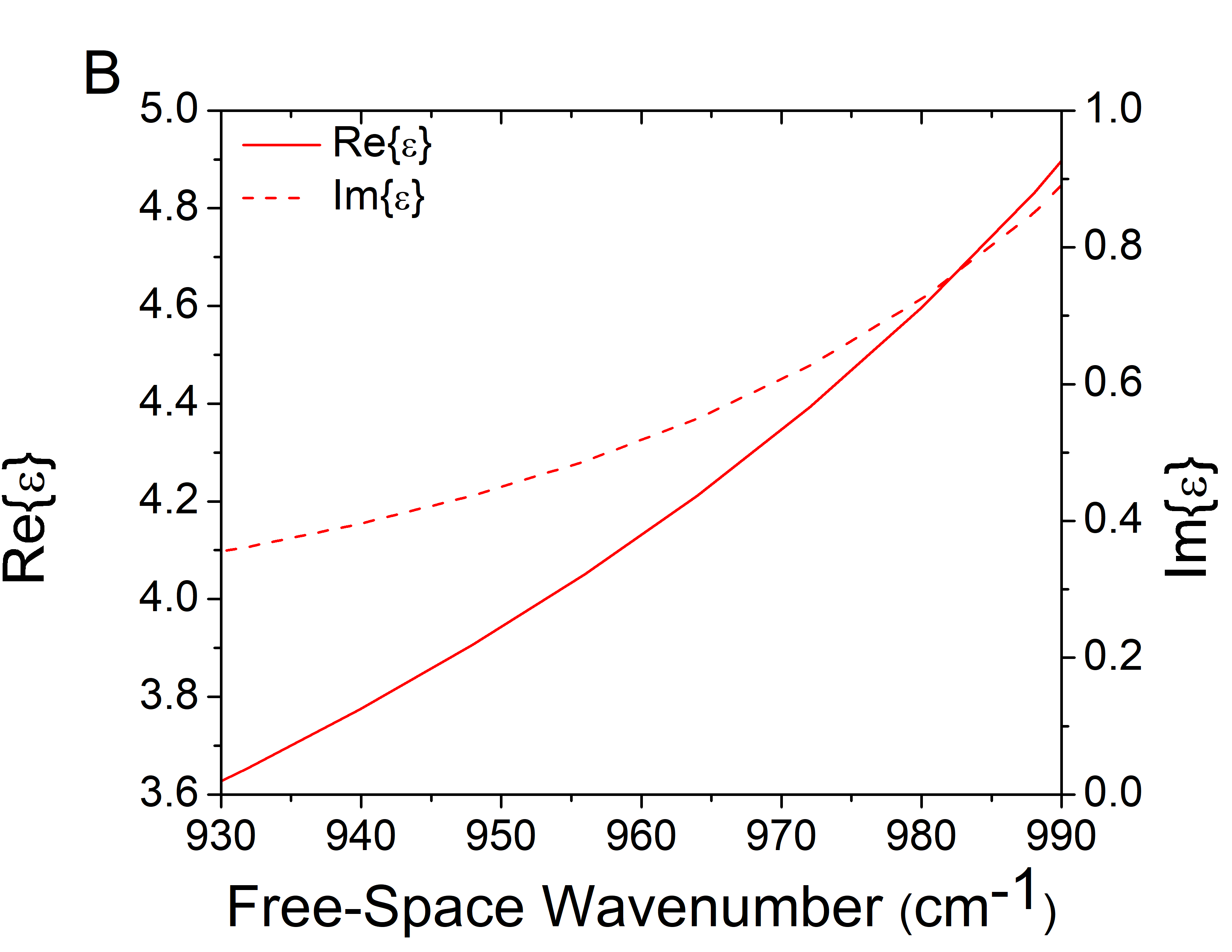}
    \caption{Dielectric functions for silicon carbide (A) and silicon dioxide (B). The data for SiC was taken from ref. \cite{dong} and for SiO$_2$ from ref. \cite{karimi}}
    \label{epsilon}
\end{figure}

\section{Methods}

To obtain the graphene plasmon dispersion, we used the condition for SPP formation\cite{abajoch}: 
\begin{equation}
    \frac{\varepsilon_1}{\sqrt{\beta^2-k_0^2\varepsilon_1}} +\frac{\varepsilon_2}{\sqrt{\beta^2-k_0^2\varepsilon_2}} + \frac{i\sigma}{\omega\epsilon_0}=0,
    \label{disp}
\end{equation}
where $\beta$ is the plasmon wavenumber, $k_0$ is the free-space radiation wavenumber, $\sigma$ is the graphene optical conductivity, $\varepsilon_1$, $\varepsilon_2$ are the dielectric functions of the superstrate, substrate, respectively, and $\epsilon_0$ is the vacuum permittivity. 

We assume graphene to be on the interface between air ($\varepsilon_1$ = 1) and either SiC or SiO$_2$. We use the Drude model for the graphene conductivity\cite{abajoch, livro-nuno}:
\begin{equation}
    \sigma(\omega) = \frac{e^2E_F\tau}{\hbar^2\sqrt{\pi}(1-i\tau\omega)},
\end{equation}
where $e$ is the electron charge, $E_F$ is the Fermi energy and $\tau$ is the relaxation rate. 
This is a good approximation when the temperature is low and electronic density is high. Also, non-local effects are neglected. These are considered important only for small nanostructures and when graphene is near metallic surfaces\cite{nature-nuno}.
The value of the relaxation time is assumed to be such that $\hbar\tau^{-1} = 2.4$ meV\cite{fei1} . 
The dielectric function for SiO$_2$ was taken to be given by\cite{karimi}.
\begin{equation}
    \varepsilon_{SiO_2} (\omega) = \varepsilon_{\infty,SiO_2} \sum_{j}\frac{s_j^2}{\omega_j^2 - \omega^2 - i\Gamma_j \omega_j},
    \label{esio2}
\end{equation}

with $\varepsilon_{\infty,SiO_2}$ = 2.3 and using the parameters shown in Table \ref{psio2}. The dielectric function of SiC was taken to be given by\cite{dong}
\begin{equation}
    \varepsilon_{SiC} (\omega) = \varepsilon_{\infty,SiC} \frac{\omega^2 - \omega^2_l + i\Gamma \omega}{\omega^2 -\omega_t^2 + i\Gamma \omega},
\end{equation}

with $\varepsilon_{\infty,SiC} = 6.7$, $\omega_l$ = 120.5 meV, $\omega_t$ = 98403 meV and $\Gamma$ = 5.9 meV.

\begin{table}
\begin{center}
\begin{tabular}{lll}
$\omega_j$ (meV) & $s_j^2$ (meV$^2$) & $\Gamma_j$(meV) \\ \hline
142            & 812           & 7.4           \\
133            & 7832          & 5.4           \\
100            & 537           & 4             \\
57             & 3226          & 6.2           \\
47             & 1069          & 24.5         
\label{psio2}
\end{tabular}
\end{center}
\caption{ parameters used for calculating the dielectric function of SiO$_2$ using equation \ref{esio2}}
\end{table}

We numerically solved equation \ref{disp} for obtaining the in-plane wavevector $\beta$, from which we can calculate both the polariton wavelength ($\lambda_{sp} = 2\pi/Re\{ \beta \}$) and the effective propagation length ($L_{eff} = 1/2Im\{\beta\}$). To better understand the plasmon-phonon interaction, we also calculated the loss function, given by the imaginary part of the p-polarized reflectivity r$_p$\cite{livro-nuno}:

\begin{equation}
    r_p = \frac{k_1\varepsilon_2 - k_2\varepsilon_1 + k_1k_2\sigma/(\varepsilon_0\omega)}{k_1\varepsilon_2 + k_2\varepsilon_1 + k_1k_2\sigma/(\epsilon_0\omega)},
\end{equation}

where $k_1$ and $k_2$ are the wavevectors of light propagating in air and the substrate, respectively.

We have also calculated the reflectance and absorbance spectra for an array of graphene ribbons in order to determine the corresponding LSP resonance condition. Since the conductivity of a regular array of graphene nanoribbons is spatially periodic ($\sigma(x+R) = \sigma(x)$, where $R$ is the periodicity), the reflected and transmitted fields can be  expressed by a Fourier-Floquet series.  Matching the boundary conditions at the interface, we arrive at the linear system of equations: \cite{livro-nuno}

\begin{align}
\left(\frac{\varepsilon_1}{\kappa_{1,n}} + \frac{\varepsilon_2}{\kappa_{2,n}}\right)E^{(1)}_{x,n} +
\nonumber \\
\frac{i}{\omega\varepsilon_0}\sum_l\tilde{\sigma}_{n-l}E^{(1)}_{x,l}=i2\frac{\varepsilon_2}{k_z}E^{\text{inc}}_x\delta_{n,0}
\end{align}

\noindent where $\beta = |k|\sin{\theta}$, $k_z = |k|\cos{\theta}$, $\kappa_{j,n} = \sqrt{(\beta + nG)^2 - \varepsilon_j\omega^2/c^2}$ and $G = 2\pi/R$. 
The Kronecker delta signals that the incident field $E^{\text{inc}}_x$ only appears for the diffraction order $n=0$. The Fourier coefficients of the conductivity $\tilde{\sigma}_l = \int_0^R\sigma e^{-iGlx}dx/R$ are calculated assuming that the width of the graphene nanoribbon in the unit cell is $d_g \le R$, and has the same conductivity  of a graphene sheet ($\sigma$). Once, the fields are obtained the reflectance, transmittance and absorbance are respectively calculated as:

\begin{align}
    \mathcal{T} = \sum_n\frac{\varepsilon_1 k_z}{\varepsilon_2|\kappa_{1,n}|}\left| \frac{E^{(1)}{x,n}}{E^{\text{inc}}_x}\right|^2 \\
    \mathcal{R} = \sum_n\frac{k_z}{|\kappa_{2,n}|}\left| \frac{E^{(2)}{x,n}}{E^{\text{inc}}_x}\right|^2 \\
    \mathcal{A} = 1 - \mathcal{T} - \mathcal{R}
\end{align}

\section{Results and discussion}

Figures \ref{epsilon}A and B show the dielectric functions of SiC and SiO$_2$, respectively. For silicon carbide the real part of the dielectric function is equal to zero at 970 cm$^{-1}$. At 951 cm$^{-1}$ the real dielectric function is -1 and the corresponding imaginary part is $< 0.04$, meaning that polaritonic losses are expected to be low. In the same spectral region, SiO$_2$ has a real dielectric function of 4 to 5, with a much higher immaginary part of around 0.4 to 0.8, meaning that the graphene plasmon wavelength is expected to be shorter and propagation losses are expected to be higher. 

\begin{figure}[t]
    \centering
    \includegraphics[scale=0.08]{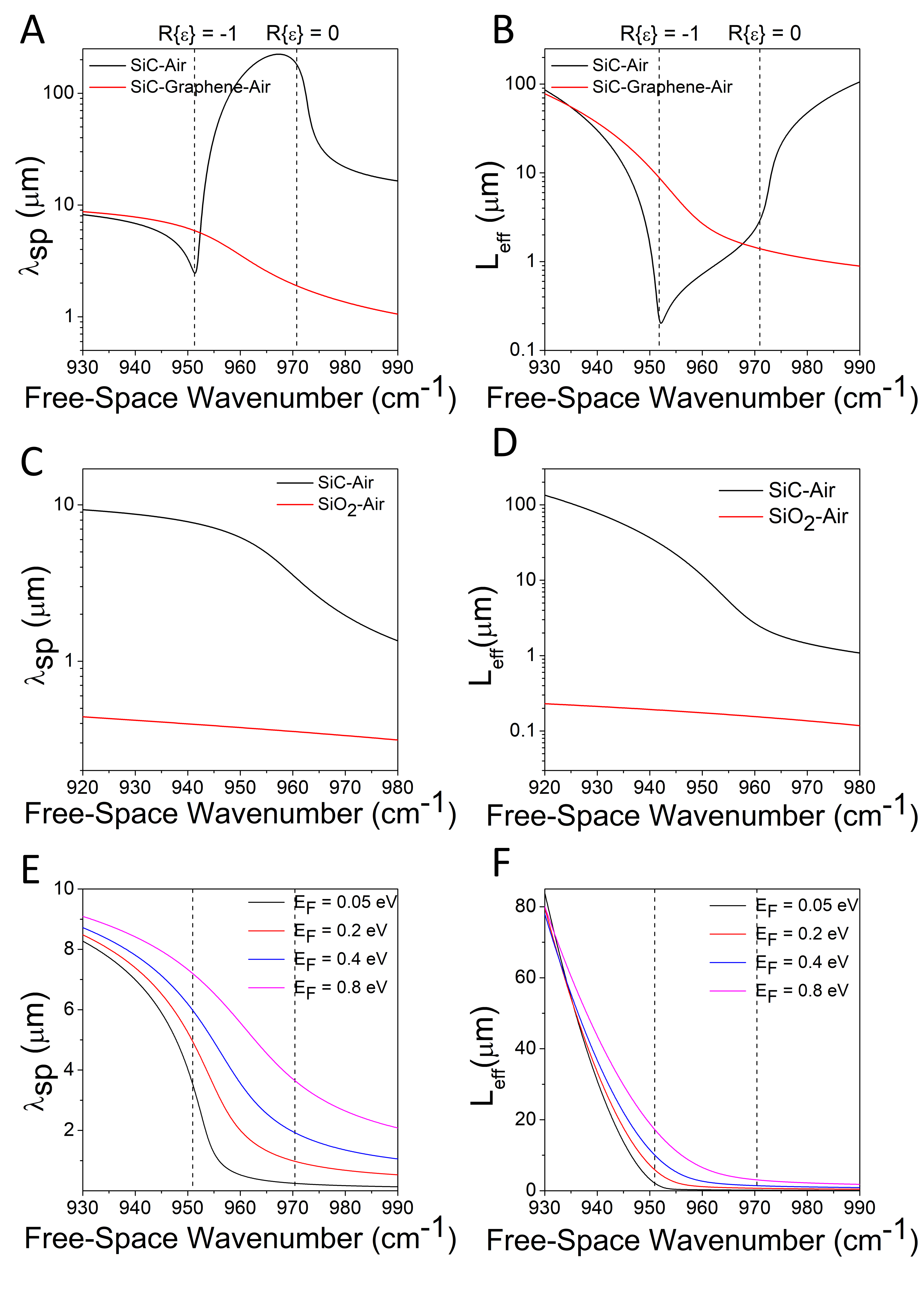}
    \caption{Polariton wavelength (A) and effective propagation length (B) on an air-SiC interface with (red line) and without (black line) graphene. Polariton wavelength (C) and propagation length (D) for graphene on SiO${_2}$ (black) and SiC (red).Figures A-D consider graphene with a 0.4 eV doping. Polariton wavelength (E) and propagation loss (F) for graphene on a SiC-air interface for various doping levels.} 
    \label{disp1}
\end{figure}

Figures \ref{disp1}A and B respectively show the polariton wavelength and effective propagation length as functions of free-space wavenumber calculated by solving eq. \ref{disp} for an air/SiC interface with and without graphene with a Fermi energy of 0.4 eV. We focus this analysis in the spectral region where the real dielectric function of the substrate is between 0 and -1. Without graphene this system should not, in principle, exhibit surface phonon polaritons for an excitation wavenumber beyond 951 cm$^{-1}$, as $Re[\varepsilon_1+\varepsilon_2]>0$, not allowing for polaritons. However, because of the imaginary part of $\varepsilon_2$, a mode arises with very high losses between 951 and 970 cm$^{-1}$, where the sum of the real dielectric functions is close to zero. The addition of graphene to the interface allows for the excitation of lower loss plasmon-phonon polaritons above 951 cm$^{-1}$, which is expressed by a shorter wavelength and a longer propagation length. For lower wavenumbers, within the Reststrahlen band, graphene imposes only a minor change to the wavelength, which suggests that in that region the polariton would exhibit a dominant SPhP character.

Figures \ref{disp1}C and D show a comparison of the polariton wavelength and propagation length using SiC and SiO$_2$ as the substrate. At the spectral range where the real dielectric function of SiC lies between -1 and 0, an order of magnitude increase in polariton wavelength is obtained. At  951 cm$^{-1}$, for example, the plasmon wavelength is 0.4 $\mu$m on SiO$_2$ and 5.8 $\mu$m on SiC. As for propagation length, it is 0.2 $\mu$m for SiO$_2$ and 9 $\mu$m for SiC. Note that for SiO$_2$, the propagation length is shorter than the polariton wavelentgh, while for SiC, the propagation length is longer. The combination of longer wavelength and propagation lengths makes polaritonic devices based on graphene on SiC simultaneously easier to fabricate and more efficient. Note that despite the polariton wavelength increase, it remains about half the light wavelength at the same frequency, allowing for subdiffractional devices to be designed. The optimal wavelength for a polaritonic device can be tailored by changing the free-space wavenumber or adjusting graphene's Fermi level.

Figures \ref{disp1}E and F respectively show the polariton wavelength and propagation length for various graphene Fermi energies, ranging from 0.05 eV to 0.8 eV. As expected, $\lambda_{sp}$ increases with E$_F$. At 951 cm$^{-1}$, for example, Fermi energies of 0.2, 0.4 and 0.8 eV result in plasmon wavelengths of approximately 4.9, 6 and 7.2 $\mu$m, respectively. 
Once again we notice that the propagation length also increases. For the same Fermi energies, we get propagation lengths of 5.8, 10 and 17.3 $\mu$m. Even for a low graphene doping such as 0.2 eV we get a much longer polariton wavelength and propagation length, on the order of micrometers, than with a SiO$_2$ substrate, which results in wavelengths of hundreds of nanometers, with much higher losses. 

\begin{figure}[t]
    \centering
    \includegraphics[scale=0.3]{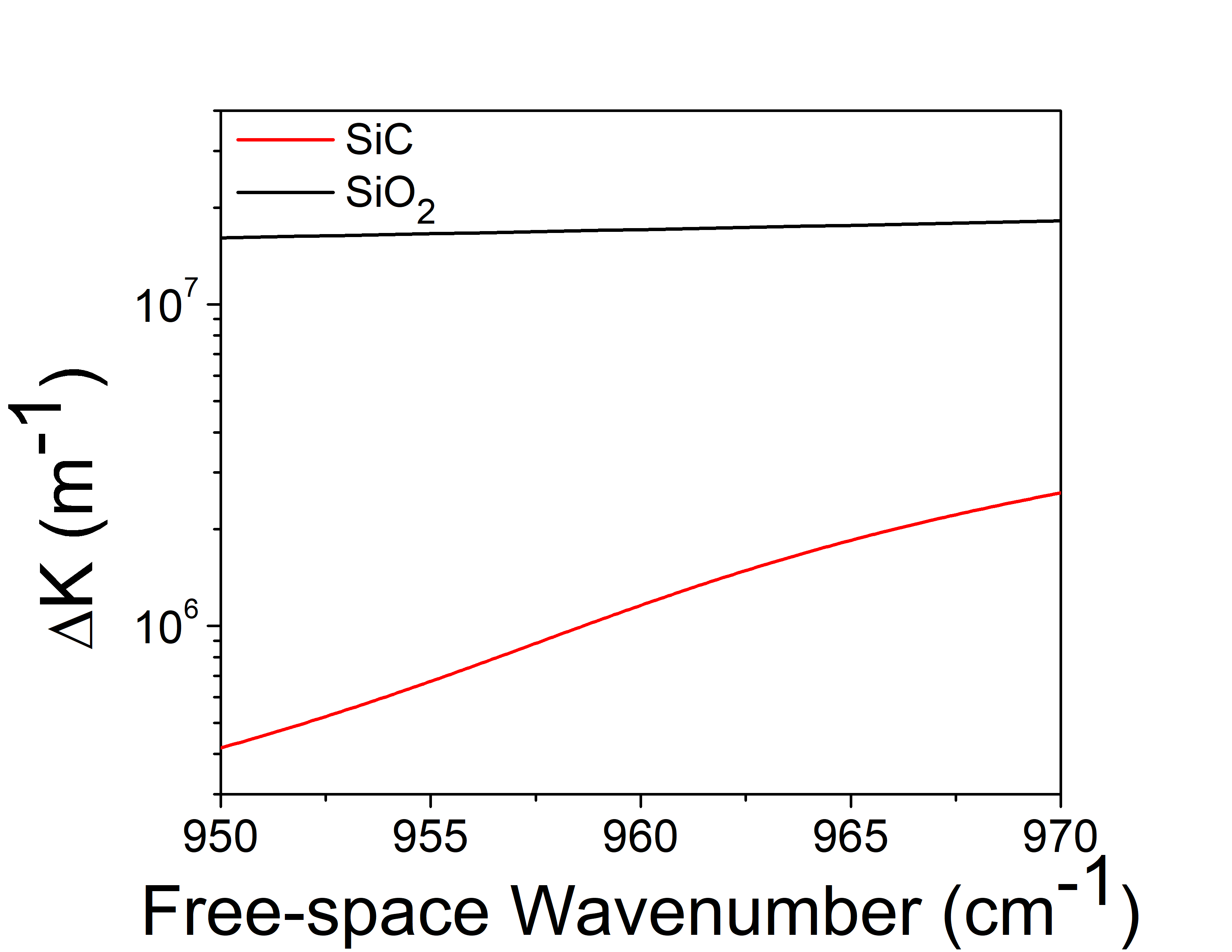}
    \caption{Phase mismatch between free-space propagating light and surface polariton for a graphene sheet at a SiC-air (red line) and a SiO$_2$-air (black line) interface with a 0.4 eV Fermi energy.}
    \label{deltak}. 
\end{figure}

In order to easily visualize the wavevector mismatch between free-space radiation and the polariton wave in graphene on the two different substrates, we plot, in Figure \ref{deltak}, $\Delta K =  |\frac{2\pi}{\lambda_0} - Re\{ \beta \}| $, where $\lambda_0$ is the free-space light wavelength, for both SiC and SiO$_2$ as substrates. 
We consider a graphene with 0.4 eV Fermi energy. A decrease of more than an order of magnitude in $\Delta$K is obtained by using SiC.
The lower wavevector mismatch between polariton and free-space radiation makes the fabrication of polaritonic devices simple. Even if we consider a graphene with a low doping of 0.2 eV, as our results show, the polariton wavelength is 4.9 $\mu$m, meaning an excitation grating, for example, with a periodicity of 2.45 $\mu$m (considering it to be half the wavelength). Such a feature size is easily fabricated with simple photolitography, rather than electron beam lithography, commonly used for conventional graphene plasmonic devices.

\begin{figure}[t]
    \centering
    \includegraphics[scale=0.14]{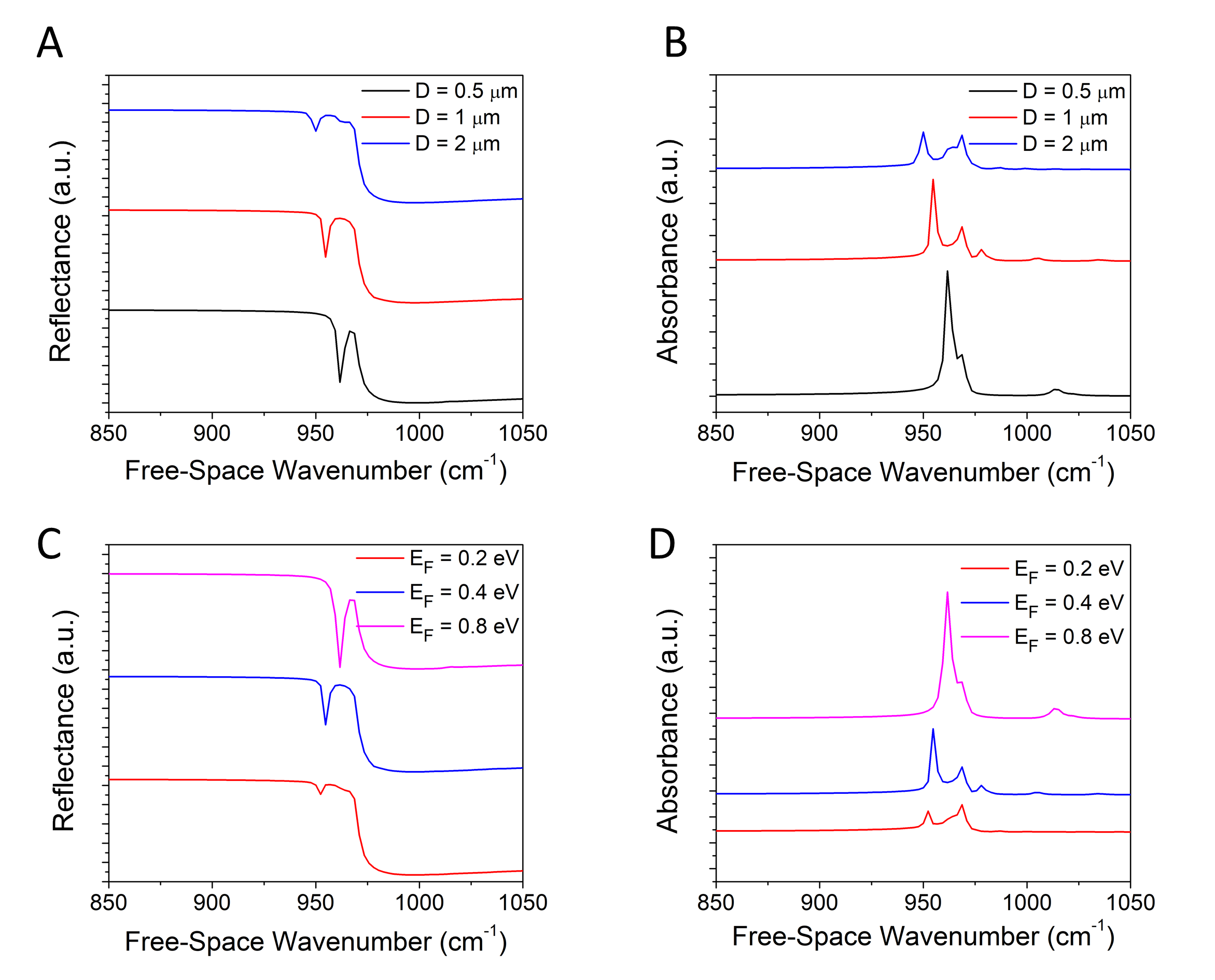}
    \caption{Polariton resonance calculations for graphene at an SiC-air interface. Reflectance (B) and absorbance (C) spectra for different graphene ribbon widths considering 0.4 eV Fermi energy. Reflectance (D) and absorbance (E) spectra for different Fermi energies considering 1$\mu$m wide graphene ribbons. }
    \label{ribbon}
\end{figure}

Since patterned graphene is one of the most convenient ways to build plasmonic devices, we now analyze the case of localized surface plasmons in microstructured graphene. We consider ribbon widths, $d_g$, from 0.5 to 2 µm, and fix the ribbon array periodicity at $R=2D$. 
Figures \ref{ribbon}A and B show the reflectance and absorbance spectra for various ribbon widths with a 0.4 eV Fermi energy. 
Since silicon carbide has a negative dielectric function below 970 cm$^{-1}$, for lower wavenumbers the material has a high reflectance, which is seen as a step in reflectance spectra. The polaritonic resonance is observed as a localized dip in reflectance and as peaks in absorbance. The resonance point can be tuned by adjusting the ribbon width and for excitation at 951 cm$^{-1}$ 2 $\mu$m ribbons are required. With $d_g= 1.0$ and 0.5 $\mu$m the polariton resonance is at 951 and 964 cm$^{-1}$, respectively. 
Figures \ref{ribbon}C and D show the same calculations for different Fermi energies considering 1 $\mu$m wide ribbons. Fermi energies of 0.2, 0.4 and 0.8 eV present polaritonic resonances at 952, 954 and 961 cm$^{-1}$ respectively.
Our calculations, therefore, show that the resonances in this spectral region can be excited using structures on the scale of micrometers, which alleviate the strain on fabrication. The absorbance peak that remains still ($\sim 970~\text{cm}^{-1}$) while   the doping level and  period of the graphene array are modified is produced by  bulk evanescent waves in the substrate (SiC). It is important to notice  that this peak is present even without the excitation of the surface polaritons.\cite{C8NR01706A} 

\begin{figure}[t]
    \centering
   \includegraphics[scale=0.17]{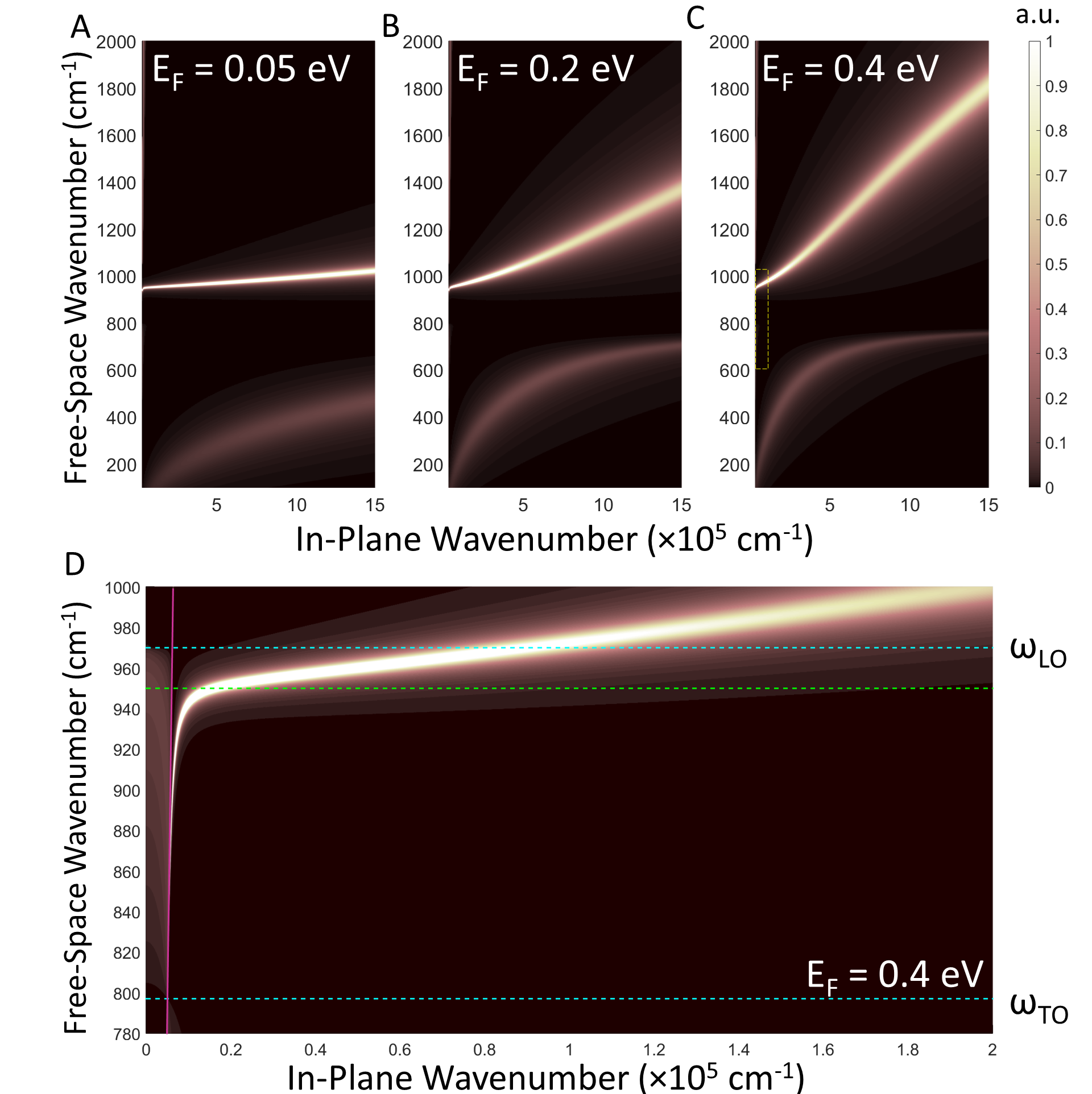}
    \caption{Loss function for graphene on a SiC/air interface with doping levels of 0.05 eV (A), 0.2 eV (B) and 0.4 eV (C). (D) zoom of (C) in the area indicated by the yellow square. The dashed cyan lines indicate the wavenumbers of SiC's longitudinal and transversal optical phonons. The dashed green line represents the wavenumber at which the real dielectric function is equal to -1, which is also the SiC's SPhP frequency. The magenta line represents the light line in SiC.}
    \label{rp}
\end{figure}

Figure \ref{rp} shows the calculated loss function for graphene at an air-SiC interface at various doping levels, through which the polariton dispersion relation can be visualized as bright yellow/white lines. 
Figures \ref{rp}A-C show the polariton dispersion for graphene with a Fermi energy of 0.05, 0.2 and 0.4 eV, respectively. If there was no graphene at the interface, the only polariton solution would correspont to a straight horizontal line at the SiC's SPhP frequency (represented by the dashed green line at \ref{rp}D).  When graphene is added to the interface, plasmon-phonon polariton hibridization takes place upon an anti-crossing of the dispersion relations, yielding 2 polariton branches, which shift towards higher free-space wavenumbers as the graphene Fermi level increases. As a general rule, the hybridized modes behave more like an SPP for low free-space wavenumbers and more like an SPhP for higher free-space wavenumbers\cite{livro-nuno}. 
Figure \ref{rp}D shows a zoom of Figure \ref{rp}C at the area indicated by the yellow rectangle. The dashed cyan lines represent the wavenumbers of the LO and TO SiC phonons, the dashed green line represents the point where Re$\{\varepsilon\}$ = -1, which is also the point where SPhP would appear for SiC without graphene and the magenta line represents light propagation through silicon carbide. The long polariton wavelengths reported here correspond to the upper hibrid polariton branch. 

It should be noted that similar hybridization takes place near the optical phonon wavenumbers of other polar dielectrics used as substrates. In fact, the same effect can be observed in SiO$_2$ at 1184 cm$^{-1}$, where its dielectric function equals -1. However, we calculate that the polariton wavelength reaches a more moderate value of 1.2 $\mu$m, with a corresponding sub-wavelength propagation length of 0.9 $\mu$m. When compared with SiC, the shorter polariton wavelength and propagation lengths are a consequence of SiO$_2$'s imaginary dielectric function, which remains high, 0.6 at the 1184 cm$^{-1}$, at the epsilon near zero point. Therefore we can conclude that, although polariton wavelength increase can be achieved for other polar dielectrics, SiC is a more efficient substrate due to its lower losses.

\section{Conclusions}

We have demonstrated a method for increasing the wavelength of graphene polaritons in order to relax the conditions for matching their wavevector with that of free-space radiation. For this purpose, we choose a substrate with a real dielectric function between -1 and 0 in the mid-infrared region: silicon carbide. In the 951 to 970 cm$^{-1}$ spectral range, we observe an order of magnitude polariton wavalength increase relative to SiO$_2$ as a substrate, which is accompanied by an order of magnitude increase in propagation length. Due to the Reststrahlen band formed by phonons in the crystal, graphene plasmons hybridize with SiC phonons, allowing for the wavelength increase, which makes the excitation of polaritons with launching gratings easier.
Along the same lines graphene structures of up to 2 $\mu$m can be used for localized surface plasmon excitation.
These result confirm SiC as a convenient substrate for graphene plasmonics, allowing for simple and large scale microfabrication methods to be employed, which can contribute to the more widespread use of graphene-based subdifractional photonics in real world applications.

\section{Acknowledgements}
This work was funded by FAPESP (grant nos. 2015/11779-4, 2018/07276-5 and 2018/25339-4), the Brazilian Nanocarbon Institute of Science and Technology (INCT/Nanocarbon), CAPES-PRINT (grant no. 88887.310281/2018-00) and CAPES-PROSUC.  NMRP acknowledges support by the Portuguese Foundation for Science and Technology (FCT) in the framework of the
Strategic Funding UIDB/04650/2020, support from the European Commission through the project ``Graphene-Driven Revolutions in ICT and Beyond'' (Ref. No. 881603, CORE 3), COMPETE 2020, PORTUGAL 2020, FEDER and
the FCT through project POCI-01-0145-FEDER-028114.

\bibliographystyle{unsrt}
\bibliography{library}
\end{document}